\documentclass[%
 reprint,
 amsmath,amssymb,superscriptaddress,floatfix,
pra,
]{revtex4-2}
\usepackage{CJK}
\usepackage{amsmath}
\usepackage{mathtools}
\usepackage{graphicx}  
\usepackage{physics}
\usepackage{graphicx}
\usepackage{dcolumn}
\usepackage{bm}

\usepackage{amssymb}
\usepackage{epstopdf}
\usepackage{inputenc}
\usepackage{esint}

\usepackage{autobreak}
 
\newcommand\V[1]{\boldsymbol{#1}} 

\begin{document}

\title{A rigorous analysis of optical forces in dielectric structures based on the Minkowski-Helmholtz formula}

\author{Huizhong Ren}
\author{Haokun Luo}
\author{Mahmoud A. Selim}
\author{Georgios G. Pyrialakos}
\author{Fan O. Wu}%
\affiliation{CREOL, The College of Optics and Photonics, University of Central Florida, Orlando, Florida 32816-2700, USA}

\author{Mercedeh Khajavikhan}%
\affiliation{Ming Hsieh Department of Electrical and Computer Engineering, University of Southern California, Los Angeles, California 90089, USA}

\author{Demetrios Christodoulides}%
\email{demetri@creol.ucf.edu}
\affiliation{CREOL, The College of Optics and Photonics, University of Central Florida, Orlando, Florida 32816-2700, USA}

\begin{abstract}
Optical forces in dielectric structures are typically analyzed by utilizing either the Maxwell stress tensor or through energy-based methods from where they can be derived by means of the eigenfrequencies and the effective refractive indices involved.While the equivalence of these two methods has been discussed in several studies, it would seem that a general electrodynamic proof of this aspect is still lacking. In this work, we provide a rigorous electrodynamic derivation based on the Minkowski-Helmholtz formula and the electromagnetic variation theorem, from where one can directly conclude that under Hermitian conditions these two approaches are formally equivalent to each other. The results of our study universally apply to any dielectric waveguide or cavity configuration. In addition, this methodology can be employed in graded index systems that do not exhibit sharp interfaces. Importantly, our analysis offers a straightforward route in predicting optical forces in a variety of photonic arrangements including dielectric scatterers and multi-element array configurations.

\end{abstract}

\maketitle

\section{\label{sec:level1}INTRODUCTION\protect\\ }
Electrodynamic forces exerted on or among dielectric structures are manifested in a ubiquitous manner in many and diverse photonic arrangements. Such forces can readily arise in a variety of optical environments like optical gradient and scattering forces on dielectric scatterers \cite{PhysRevLett.24.156,Ashkin:86,PhysRevLett.54.1245,ashkin1971optical}
and forces induced between evanescently coupled waveguides and cavities \cite{li2009tunable,li2008harnessing} (Fig.1).These same electromagnetic forces are also at play in settings where the refractive index can vary gradually in space, as for example, in graded-index fibers and liquids\cite{ashkin1973radiation,brasselet2008liquid}. During the last two decades or so, the electromagnetic forces between two dielectric elements (cavities or waveguides) have been theoretically analyzed by relying mainly on the following two approaches: (a) the use of the Maxwell stress tensor formalism \cite{jackson1999classical} and (b) an energy-based method from where one can extract the force among two elements from the spatial gradient of the respective eigenvalues \cite{povinelli2005evanescent,povinelli2005high,ma2011mechanical,rakich2011scaling,rakich2009general,rakich2010tailoring,rodrigues2017optical,rodrigues2017rigorous,rodrigues2019geometric}. In this regard, the bonding and anti-bonding forces between two waveguides were first investigated by Povinelli \textit{et al} \cite{povinelli2005evanescent} where it was found numerically that these two methodologies are indeed consistent with each other. In this pioneering study, the electromagnetic problem was theoretically addressed by effectively embedding the two-core waveguide system under consideration within a virtual optical cavity, from where the forces can be evaluated through the variation of the corresponding eigenfrequencies.  In this same spirit, the same problem has been systematically studied in subsequent works using the response theory of optical forces, transformation optics schemes, and numerical simulations \cite{povinelli2005high,ma2011mechanical,rakich2011scaling,rakich2009general,rakich2010tailoring,rodrigues2017optical,rodrigues2017rigorous,rodrigues2019geometric,iizuka2019modal,pernice2009theoretical}. At this juncture, the following question arises. Given that waveguides are broadband systems and hence by nature lack eigenfrequencies, to what extent will such a hybrid treatment (involving a virtual cavity) is indeed applicable, and if so, how does it formally reconcile with the Maxwell stress tensor? Quite recently, this assertion was proved for 1D planar waveguides using the Hellmann-Feynman theorem \cite{miri2018optical}. Yet, at this point, it would seem that a general and formal electrodynamic proof of the aforementioned equivalence is still lacking. 
\begin{figure}
\includegraphics[width=8.6cm]{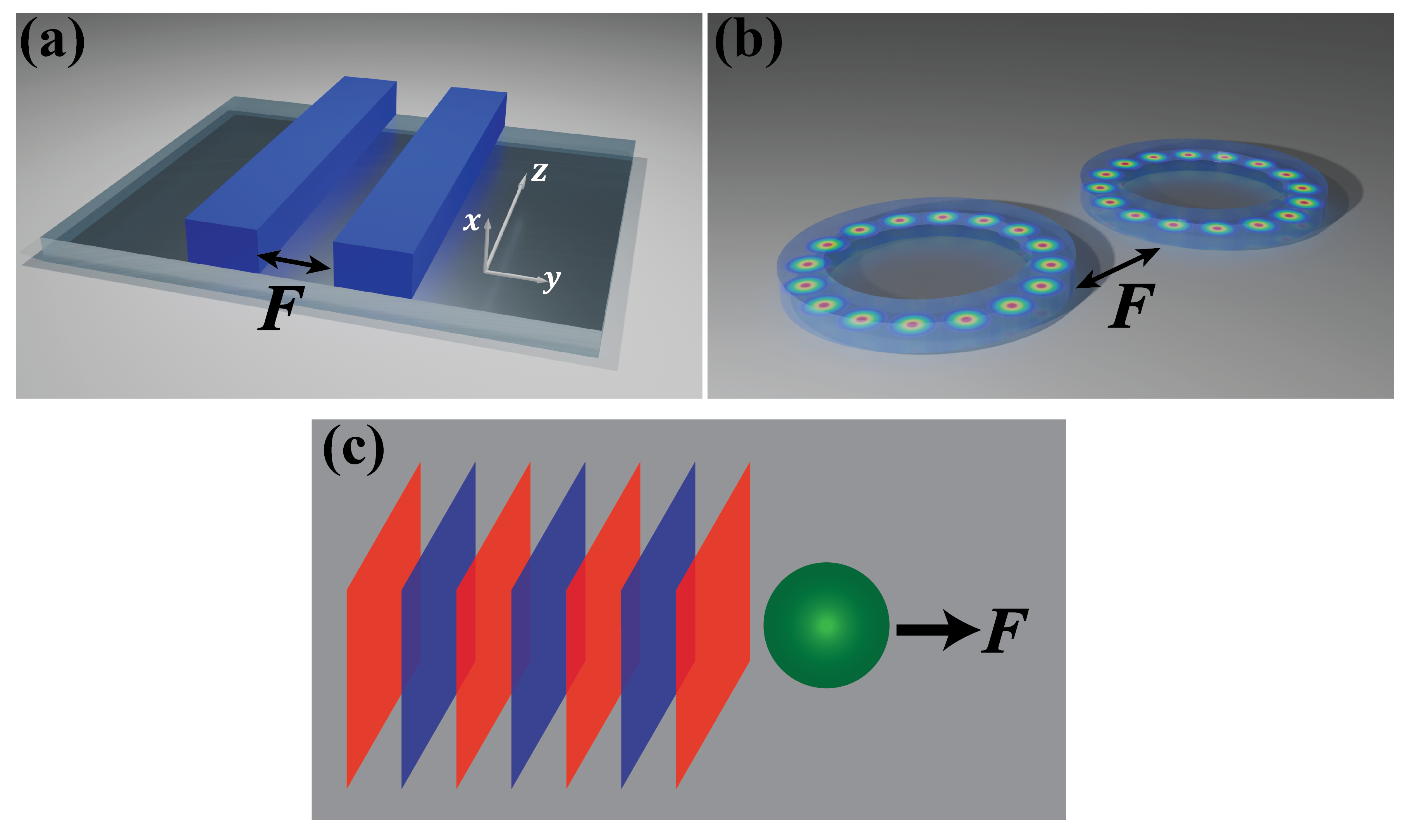}
\caption{\label{fig:1}Optical forces are exerted between two evanescently coupled (a) dielectric waveguides and (b) optical cavities. (c). Optical force acting on a dielectric spherical scatterer induced by a plane wave.}
\end{figure}

In this article, we rigorously prove that under Hermitian conditions, the energy-based method \cite{povinelli2005evanescent} is fully equivalent to the Maxwell stress tensor formalism. This is made possible by employing the Minkowski-Helmholtz formula when used in conjunction with the electromagnetic variation theorem-a byproduct of the Lorentz reciprocity theorem. Our theoretical results are general and therefore applicable to any arbitrary dielectric system involving optical cavities and waveguides. In addition, the Minkowski-Helmholtz formalism can be readily deployed in analyzing more complex arrangements like optical scatterers, multi-element cavities and waveguide arrays, as well as graded-index guiding elements. Finally, this same approach can serve as a powerful tool by means of which one can intuitively understand the way optical forces act in complex photonic settings, that go beyond two-element structures considered so far in the literature. Numerical simulations corroborate our theoretical analysis.

\section{Theoretical analysis of the induced optical forces}
We begin this work by invoking the Helmholtz formula in electrodynamics which provides an alternative route in analyzing the induced force density in a material with constitutive parameters $\epsilon (\V{r})=\epsilon_0 \epsilon_r (\V{r})$, $\mu (\V{r})=\mu_0 \mu_r (\V {r})$. The Helmholtz components, as derived from energetic considerations, can then be recast in the so-called Minkowski force density $\V {{f_M}}$(force per unit volume), which on average, for time harmonic fields is given by \cite{vago1998electromagnetic,zangwill2013modern,gordon1973radiation,stratton2007electromagnetic}
\begin{equation}
\langle{\V{f_M}}\rangle=\frac{1}{2}Re\left(\rho_f \V{E^*}+\V{j_f}\times\V{B^*} -\frac{1}{2}|\V{E}|^2\nabla\epsilon-\frac{1}{2}|\V{H}|^2\nabla\mu\right).
\end{equation}

In the Minkowski-Helmholtz formula (Eq. (1)), $\rho_f$, $\V{j_f}$ represent the free electric charge and current density, respectively, while $\V{E}$,$\V{H}$ denote the time harmonic electric and magnetic fields. Meanwhile, $|\V{E}|^2=\V{E}\cdot\V{E^*}$ and $|\V{H}|^2=\V{H}\cdot\V{H^*}$. The first two terms in Eq. (1) correspond to the Lorentz force density, whereas the last two contribute to the optical force through the inhomogeneity/discontinuity of the medium itself. From this point forward in this paper, we will assume that the dielectric system is lossless (Hermitian) and $\mu_r=1$.  For a typical dielectric non-magnetic material, and in the absence of any free currents and charges, Eq. (1) is reduced to the following simple expression
\begin{equation}
    \langle{\V{{f_M}}}\rangle=-\frac{|\V{E}|^2\nabla\epsilon}{4},
\end{equation}
which indicates that the force density results only from the inhomogeneities/discontinuities in the electric permittivity. Note that the Minkowski-Helmholtz force density is formally related to the Maxwell stress tensor via $\langle{\V{{f_M}}}\rangle=\nabla\cdot\langle{\tensor{T}}\rangle$ where $\langle{T_{ij}}\rangle=\frac{1}{2}Re\left[\epsilon E_i E_j^*+\mu H_i H_j^*-\frac{1}{2} \delta_{ij}(\epsilon |\V{E}|^2+\mu |\V{H}|^2)\right]$. Of importance will be to first understand how the Minkowski-Helmholtz force can be described in the presence of sharp boundaries or index ($\epsilon_r (\V{r})=n^2 (\V{r})$) discontinuities (Fig. 2(a)). At an abrupt interface $(n_1,n_2)$, the boundary conditions for the tangential and normal electric field components imply $\V{E}_{1,t}=\V{E}_{2,t}$ and $\V{D}_{1,n}=\V{D}_{2,n}$ (Fig. 2(b)). In this case, the sharp dielectric boundary can be described by a Heaviside step function, $\epsilon_r (\V{r})=\epsilon_{r_1} (\V{r})+[\epsilon_{r_2} (\V{r})-\epsilon_{r_1} (\V{r})]H(\V{r}-\V{r_0})$ where $\V{r_0}$ represents a position vector on the boundary. From here, the optical force acting on the infinitesimal surface $d\V{a}=da\hat{\V{n}}$ can be evaluated from the integral of $\langle\V{{f_M}}\rangle$ over a selected volume enclosed by the surface $\sigma$ as shown in Fig. 2(a). In this respect, the optical force involves two components, arising from the tangential and normal electric fields. In general, we can write $\nabla\epsilon$ along the normal unit vector of the surface $\hat{\V{n}}$ as $\nabla\epsilon=\hat{\V{n}}d\epsilon/dr=(\epsilon_2-\epsilon_1 )\delta(\V{r}-\V{r_0})\hat{\V{n}}$. Therefore, the force per unit area acting on the surface is $\V{{f_M}}={f_M}\hat{\V{n}}$ where
\begin{equation}
\begin{aligned}
    {f_M} & = -\frac{1}{4}\left (\int_{\V{r_0^-}}^{\V{r_0^+}} |\V{E}_t|^2 \frac{d\epsilon}{dr}\, dr \ +\int_{\V{r_0^-}}^{\V{r_0^+}} \frac{|\V{D}_n|^2}{\epsilon^2} \frac{d\epsilon}{dr}\, dr\right)\\
    & = -\frac{1}{4}\left[\int_{\V{r_0^-}}^{\V{r_0^+}} |\V{E}_t|^2 \frac{d\epsilon}{dr}\, dr \ -\int_{\V{r_0^-}}^{\V{r_0^+}} |\V{D}_n|^2\frac{d}{dr}\left(\frac{1}{\epsilon}\right)\, dr\right]\\
    & = \frac{1}{4}\left[\left(\epsilon_1-\epsilon_2\right)(|\V{E}_t|^2)_{\V{r}=\V{r_0}}+\left(\frac{1}{\epsilon_2}-\frac{1}{\epsilon_1}\right)(|\V{D}_n|^2)_{\V{r}=\V{r_0}}\right].
\end{aligned}
\end{equation}
The expression above is general in the sense that it can be utilized to analyze optical forces in any arbitrary photonic arrangement with sharp index discontinuities in the absence of any surface charges $(\V{D}_{1,n}=\V{D}_{2,n})$, i.e., dielectric scatterers, multi-element array configurations, cavities, etc.  
\begin{figure}
\includegraphics[width=8.6cm]{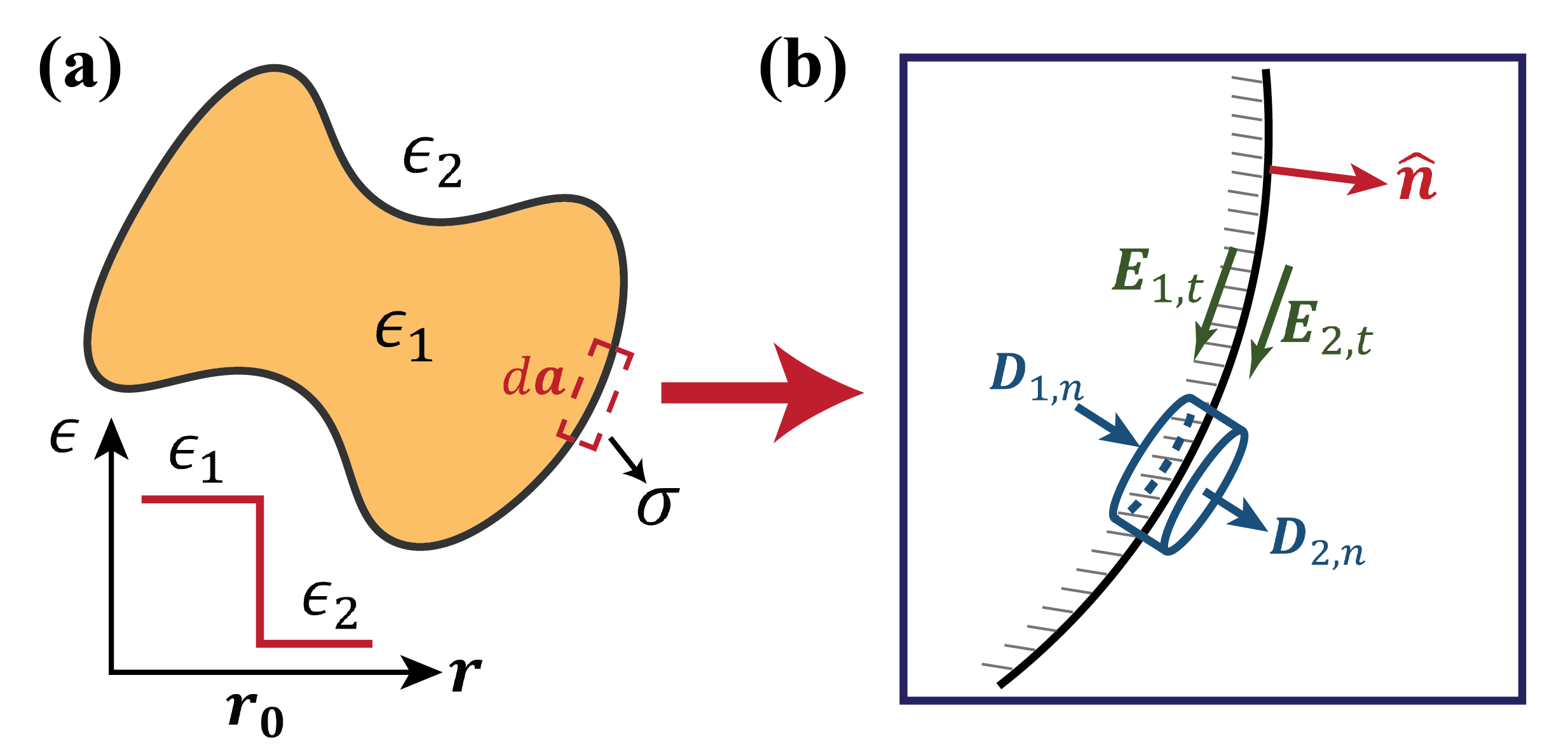}
\caption{\label{fig:2}(a) An arbitrary optical dielectric element with a sharp permittivity index discontinuity. (b) The electromagnetic field boundary conditions on the infinitesimal surface $d\V{a}$. The tangential components of the electric field $\V{E}_t$ and the normal components of the electric displacement $\V{D}_n$ are continuous.}
\end{figure}

To calculate the electrodynamic forces, we here use the principle of virtual work. In this respect, the optical force exerted on the dielectric object can be derived from the virtual work $\delta W$ produced when a virtual displacement $\delta\V{\xi}$ (as shown in Fig. 3(b)) occurs,
\begin{equation}
    \delta W=\V{F}\cdot\delta\V{\xi}.
\end{equation}
In the formalism used in this study, the virtual work $\delta W$ between say two dielectric structures (Fig. 3(a, b)), can now be directly evaluated from the Minkowski-Helmholtz force density, i.e.,
\begin{equation}
    \delta W=\delta\V\xi\cdot\iiint\langle{\V {f_M}}\rangle\,dv \ =-\frac{1}{4}\iiint|\V E|^2\nabla \epsilon\cdot\delta\V\xi\, dv \ .
\end{equation}
As shown in Fig. 3(c), by virtually displacing for example the left element, the electric permittivity profile undergoes a virtual change $\delta\epsilon$ i.e., $\delta\epsilon=\epsilon(\V r-\delta\V\xi)-\epsilon(\V r)=-\nabla\epsilon\cdot\delta\V\xi$. Therefore, one can deduce that
\begin{equation}
    \delta W=\frac{1}{4}\iiint|\V E|^2\delta\epsilon\, dv \ .
\end{equation}
\begin{figure}
\includegraphics[width=8.6cm]{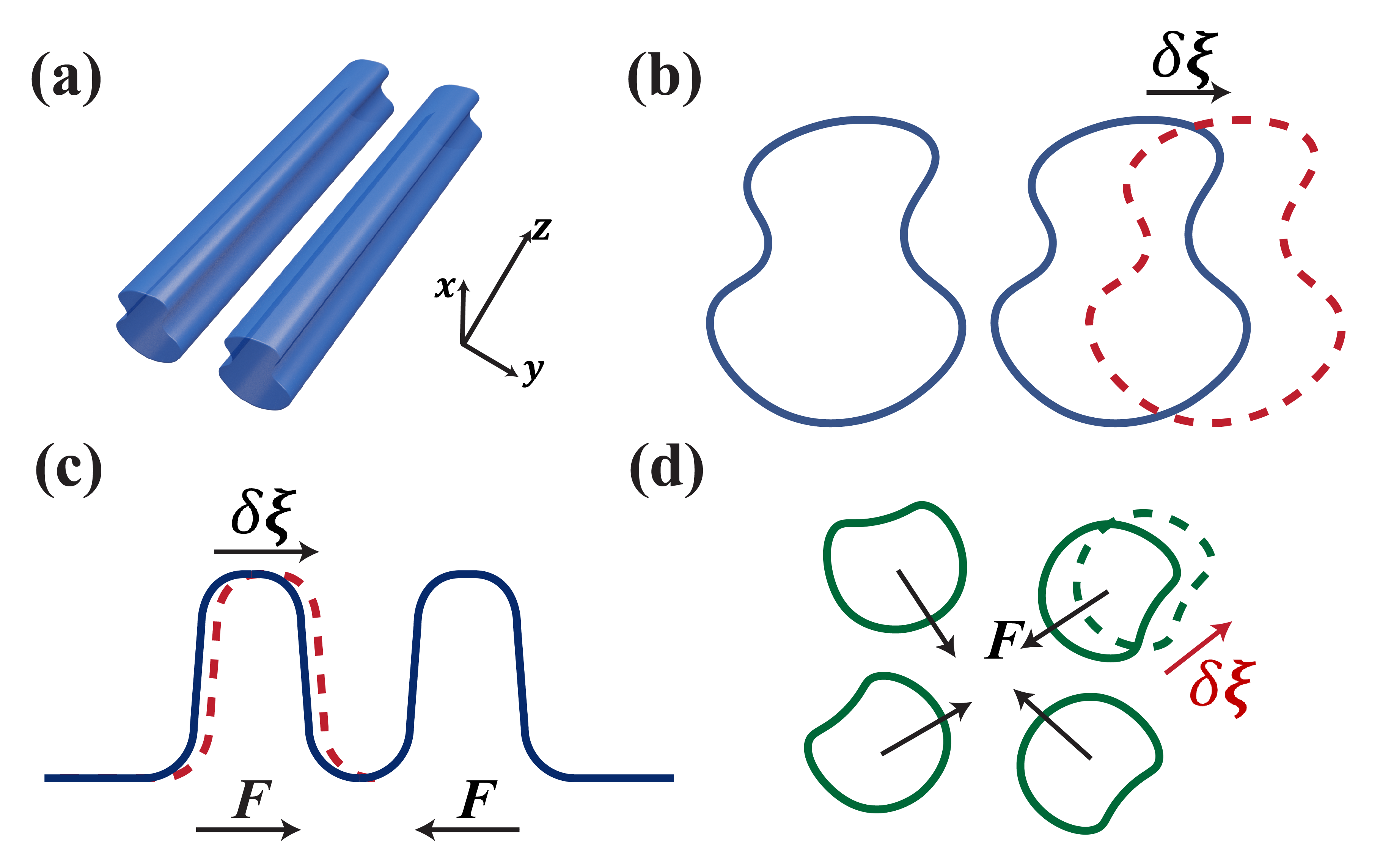}
\caption{\label{fig:3}(a) Schematic of a coupled waveguide system where radiation forces are exerted. (b) One waveguide element in (a) is virtually displaced by $\delta\V\xi$. (c) The permittivity change $\delta\epsilon$ resulting from the virtual displacement $\delta\V\xi$. The blue and red lines denote the permittivity profiles before and after this displacement. (d) Optical forces emerging in an arbitrary multi-element dielectric structure.}
\end{figure}

In order to complete this electrodynamic proof concerning the equivalence between the Maxwell stress tensor formalism and the energy-based method, we will next consider two versions of the electromagnetic variation theorem, corresponding to either waveguide systems \cite{alferness2013guided} or cavity structures \cite{waldron1960perturbation,christodoulides2002discrete}. In this regard, without any loss of generality, let us first investigate an arbitrary dielectric waveguide arrangement like that depicted in Fig. 3(a). Assuming that wave propagation takes place along the $z$ direction (Fig. 3(a)), the time-harmonic electromagnetic modes are described by $\V E(\V r)=\V {E_0} (x,y) e^{i\beta z}$ and $\V H(\V r)=\V {H_0} (x,y) e^{i\beta z}$, where $\V {E_0} (x,y)$ and $\V {H_0} (x,y)$ represent the waveguide electric/magnetic spatial eigenmodes having a propagation constant $\beta$. If a variation $\delta\epsilon$ is performed on the permittivity profile of this waveguide arrangement (because of a virtual displacement $\delta\V\xi$, as in Fig. 3(c)), then the fields are also perturbed according to $\delta\V E=(\delta \V {E_0}+iz\delta\beta\V {E_0}) e^{i\beta z}$ and $\delta \V H=(\delta \V {H_0}+iz\delta\beta\V {H_0}) e^{i\beta z}$, and hence $\nabla\times\delta\V E=i\omega\mu\delta\V H$ and $\nabla\times\delta\V H=-i\omega\delta(\epsilon\V E)$. By combining these latter expressions with the vector identity $\nabla\cdot(\V A\times\V B)=(\nabla\times\V A)\cdot\V B-\V A\cdot(\nabla\times\V B)$ and after using Maxwell’s equations, one obtains
\begin{equation}
    \nabla\cdot(\V E^*\times\delta\V H+\delta \V E\times\V H^*)=i\omega\delta\epsilon |\V E|^2.
\end{equation}
After inserting the corresponding field expressions into Eq. (7), we find,
\begin{equation}
    \nabla\cdot\left(\V {E_0}^*\times \delta\V {H_0}+\delta\V {E_0}\times\V {H_0}^*+4iz\delta\beta\V S\right)=i\omega\delta\epsilon|\V {E_0}|^2,
\end{equation}
where $\V S(x,y)=S_z \hat{\V z}=\frac{1}{2} Re(\V E\times \V H^*)=\frac{1}{4}(\V {E_0}\times \V {H_0}^*+\V {E_0}^*\times \V {H_0})$ denotes the time averaged Poynting vector. In this case, Eq. (8) can be rewritten as
\begin{equation}
    \nabla_t\cdot\V g+4i\delta\beta S_z=i\omega\delta\epsilon|\V {E_0}|^2,
\end{equation}
where $\V g=\V {E_0}^*\times\delta\V {H_0}+\delta \V{E_0}\times \V {H_0}^*+4iz\delta\beta\V S$. After integrating Eq. (9) over a cross-section $z=\textrm{const}$ of this waveguide, we find,
\begin{equation}
    \iint\left(\nabla_t\cdot\V g+4i\delta\beta S_z\right)\,dx\,dy\ =\iint i\omega\delta\epsilon|\V{E_0}|^2\,dx\,dy.
\end{equation}
By using the divergence theorem on the LHS of Eq. (10), we obtain $\iint\nabla_t\cdot\V g\,dx\,dy\ =\oint_C\V g\cdot\V {e_t}\,dl\ $, where the line integral is taken over an infinitely large contour $C$ enclosing the waveguide cross section where $\V {e_t}$ is a unit vector that is normal to the contour. Given that the modal fields $\V {E_0}$ and $\V {H_0}$ are associated with bound modes that vanish at infinity, this line integral is equal to zero. Thus, Eq. (10) can be further reduced to $4i\delta\beta P=\iint i\omega\delta\epsilon|\V {E_0}|^2\,dx\,dy$, where $P=\iint S_z\,dx\,dy$ is the time averaged power conveyed by the corresponding mode. To this end, one can obtain the change in the propagation constant $\delta\beta=k_0\delta n_{eff}$ due to this perturbation $\delta\epsilon$ from \cite{alferness2013guided},
\begin{equation}
    \delta\beta=\frac{\omega\iint \delta\epsilon|\V {E_0}|^2\,dx\,dy}{4P}.
\end{equation}
We note that $\delta\epsilon$ in Eq. (11) is again given by $\delta\epsilon=\epsilon(\V r-\delta\V\xi)-\epsilon(\V r)=-\nabla\epsilon\cdot\delta\V\xi$, and in this waveguide arrangement, the volume integral in Eq. (6) can be expressed as $\iiint dv=L\iint dxdy$ where $L$ is the length of the waveguide. In this respect, by combining Eqs. (4), (6) and (11), one obtains
\begin{equation}
    \V F\cdot\delta\V\xi=\frac{PL}{c}\delta n_{eff}.
\end{equation}
This latter expression can now be rewritten as 
\begin{equation}
    F=\frac{PL}{c} \frac{dn_{eff}}{d\xi},
\end{equation}
where $F=\V F\cdot\delta\hat{\V \xi}$ and $\delta \hat{\V\xi}$ is the unit vector associated with $\delta\V\xi$. The relation expressed in Eq. (13) is identical to that previously obtained using energy-based methods \cite{povinelli2005evanescent,povinelli2005high,rakich2009general}. This completes the proof for waveguide configurations. In essence, by utilizing the electromagnetic variation theorem in conjunction with the Minkowski-Helmholtz formula, we have rigorously shown that the Maxwell stress tensor formalism is formally equivalent to previously developed energy-based methods \cite{povinelli2005evanescent,povinelli2005high,rakich2009general}. Our result is general and applies to any multi-waveguide arrangement provided that the elements are all electromagnetically coupled (Fig. 3(d)) where $n_{eff}$ denotes the effective refractive index of a particular supermode in this configuration.

Our analysis can be readily extended to cavity set-ups \cite{waldron1960perturbation,christodoulides2002discrete}. The electromagnetic modes of a cavity resonator can be written as $\V E(\V r)=\V {E_0} (\V r) e^{-i\omega t}$ and $\V H(\V r)=\V {H_0} (\V r) e^{-i\omega t}$, where $\V {E_0} (\V r)$ and $\V {H_0} (\V r)$ represent the cavity eigenmodes while $\omega$ stands for their corresponding eigenfrequency. As before, we assume that one of the cavity elements is virtually displaced by $\delta\V\xi$, producing work $\delta W=\V F\cdot\delta\V\xi$. In turn, this virtual displacement leads to a variation in the eigenmode fields $(\delta\V {E_0}, \delta\V {H_0})$ and eigenfrequencies ($\delta\omega$) due to a change in the permittivity profile $\delta\epsilon=-\nabla\epsilon\cdot\delta\V\xi$. Electrodynamically, these variations obey $\nabla\times\delta\V {E_0}=i\delta(\omega\mu\V {H_0})=i[\V {H_0} \delta(\omega\mu)+\omega\mu\delta\V {H_0}]$ and $\nabla\times\delta\V {H_0}=-i\delta(\omega\epsilon\V {E_0})=-i[\V {E_0}\delta(\omega\epsilon)+\omega\epsilon
\delta\V {E_0}]$. As before, by using the vector identity $\nabla\cdot(\V A\times\V B)=(\nabla\times\V A)\cdot\V B-\V A\cdot(\nabla\times\V B)$ and Maxwell’s equations, we find that
\begin{equation}
\begin{aligned}
    &\nabla\cdot\left(\V {E_0}^*\times\delta\V {H_0}+\delta\V {E_0}\times\V {H_0}^*\right)\\
    &=i\left[\delta(\omega\epsilon)|\V {E_0}|^2+\delta(\omega\mu)|\V {H_0}|^2\right].
\end{aligned}
\end{equation}
By applying the divergence theorem to Eq. (14), one quickly obtains 
\begin{equation}
\begin{aligned}
    &\oiint\left(\V {E_0}^*\times\delta\V {H_0}+\delta\V {E_0}\times\V {H_0}^*\right)\cdot d\V a\\
    &=i\iiint\left[\delta(\omega\epsilon)|\V {E_0}|^2+\delta(\omega\mu)|\V {H_0}|^2\right]\,dv.
\end{aligned}
\end{equation}
Here, the surface integral extends over a virtual infinitely large-closed surface $\V a$ surrounding the cavity. Since $\V {E_0}$ and $\V {H_0}$ represent bound modes that vanish at infinity, the surface integral in Eq. (15) is zero. For a non-magnetic dielectric material, Eq. (15) can now provide the shift (variation) in the eigenfrequency $\delta\omega$ in this cavity system
\begin{equation}
    \delta\omega=-\frac{\omega\iiint\delta\epsilon|\V {E_0}|^2\,dv}{4U},
\end{equation}
where $U$ is the time-averaged energy stored in the cavity that is given by $U=\left[\iiint\left(\epsilon |\V {E_0}|^2+\mu|\V {H_0}|^2\right)\,dv\right]/4$. In the same vein, using Eqs. (4) and (6), one can obtain
\begin{equation}
    \V F\cdot\delta\V \xi=-\frac{U}{\omega}\delta\omega,
\end{equation}
or equivalently,
\begin{equation}
    F=-\frac{U}{\omega}\frac{d\omega}{d\xi},
\end{equation}
where $F=\V F\cdot\delta\hat{\V\xi}$. Eq. (18) is identical with the expression previously obtained by Povinelli \textit{et al}. through quantum arguments \cite{povinelli2005evanescent,povinelli2005high}. This now completes the proof for cavity arrangements.

\begin{figure}
\includegraphics[width=8.6cm]{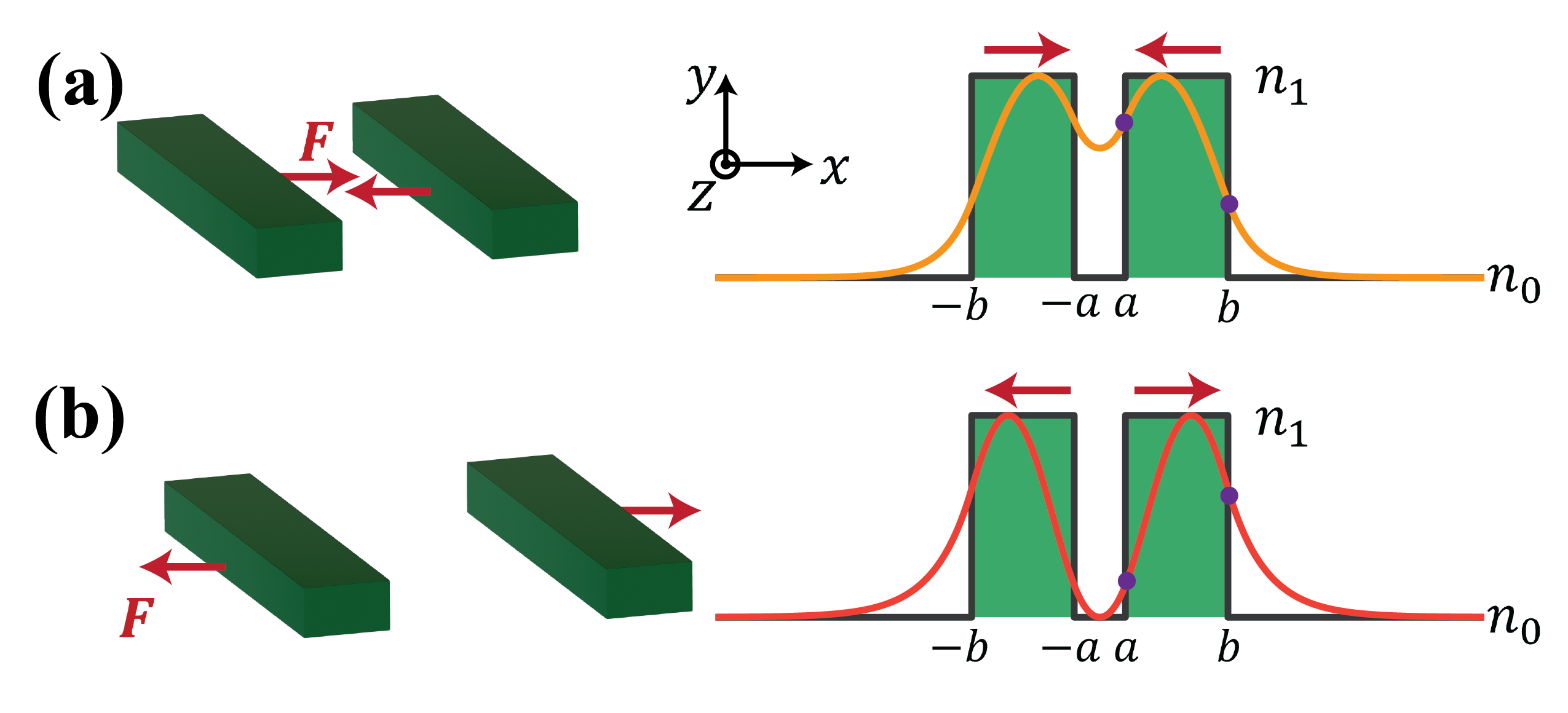}
\caption{\label{fig:4}Coupled slab waveguides and the resulting attractive and repulsive forces for the even (a) and odd (b) supermodes. The black outline represents the refractive index distribution, while the colored lines denote the mode intensity profiles.}
\end{figure}
                                                                                                   
\section{Electrodynamic forces in various complex arrangements}
It is worth emphasizing that the Minkowski-Helmholtz formula can provide an intuitive understanding as to how optical forces act on multi-element structures. To demonstrate this aspect, we will next consider a simple configuration consisting of two planar step index waveguides, as shown in Fig. 4. The arrangement is centered at the origin and the waveguides extend between $-b<x<-a$ and $a<x<b$, thus guiding optical waves along $z$ direction. The cladding refractive index is assumed to be $n_0$ while in the guiding layers is $n_1$. Here, the optical force exerted on each isolated waveguide can now be analyzed using the Minkowski-Helmholtz formula. If we consider the waveguide on the right in Fig. 4, the force per unit area is given by $\V f=\int_{a^-}^{b^+}\left(-\frac{1}{4}|\V E|^2\nabla\epsilon\right)\,dx$. From Eq. (3), for a transverse electric (TE) mode, for example, this expression is reduced to
\begin{equation}
    \V f=\frac{\epsilon_0}{4}(n_1^2-n_0^2)\left[\left.\left(|E_y|^2\right)\right\vert_{x=b}-\left.\left(|E_y|^2\right)\right\vert_{x=a}\right]\hat{\V x}.
\end{equation}
Evidently, for an even mode as depicted in Fig. 4(a), the field amplitude at the inner edge of the slab is larger than that at the outer edge, i.e., $\left.\left(|E_y|^2\right)\right\vert_{x=a}>\left.\left(|E_y|^2\right)\right\vert_{x=b}$, and hence, according to Eq. (19), this leads to an attractive force. On the other hand, for an odd mode (Fig. 4(b)), the presence of a node at the origin leads to $\left.\left(|E_y|^2\right)\right\vert_{x=a}<\left.\left(|E_y|^2\right)\right\vert_{x=b}$ which in turn results in a repulsive force. In this regard, the Minkowski-Helmholtz formula provides an intuitive tool in predicting both the magnitude and direction of the optical force exerted on each component of this coupled photonic arrangement. 
 
In general, the Minkowski-Helmholtz formula can be employed to analyze optical forces in more complex photonic arrangements that go beyond two-element structures. For example, let us consider a linear waveguide array comprising of $N$ step-index circular guiding elements, each one of them evanescently coupled to its nearest neighbors, as depicted in Fig. 5(a). Each waveguide is assumed to be single-moded, i.e., it only supports the $LP_{01}$ mode. The core radius of each element is $a$ and the distance between any two waveguides is $D$. By adopting the formalism of coupled mode theory, the supermode field distribution can be approximately expressed as 
\begin{equation}
    \psi^m(x,y)=\sum_n c_n^m G_0(x-nD,y).
\end{equation}
\begin{figure}
\includegraphics[width=8.6cm]{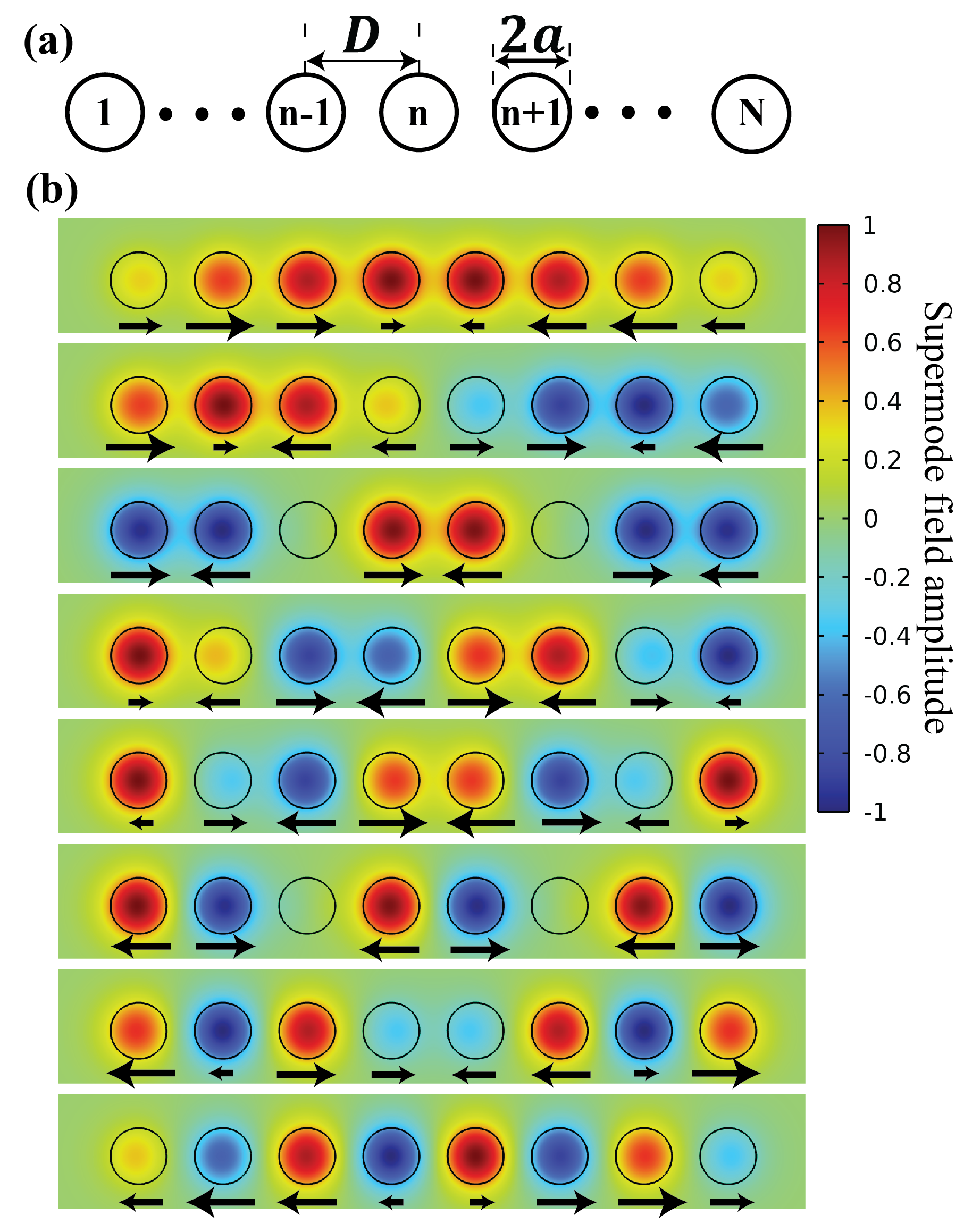}
\caption{\label{fig:5}(a) A linear array structure composed of $N$ step-index weakly guiding elements. (b) Numerical simulation results for 8 elements using finite element methods. The arrows represent the force exerted on each element, as evaluated from Eq. (22).}
\end{figure} 
In this expression, $c_n^m$ denotes the supermode field amplitude at each site, i.e., $c_n^m=\sqrt{2/(N+1)} \sin{[m\cdot n\pi/(N+1)]}$ \cite{xia2015supermodes}, where the integers $m,n=1,2,…,N$ stand for the supermode and the local site indices, respectively. Meanwhile, $G_0 (x,y)$ represents the local $LP_{01}$ mode profile in each element. Within the tight-binding approximation, the field at the $n$-th waveguide can be obtained through the superposition of the local mode profiles, that is $\phi_n^m(x,y)=\sum_{k=n-1}^{k=n+1}c_k^m G_0(x-kD,y)$. For simplicity, we now place the origin of the coordinate system at the center of the selected waveguide. From the Minkowski-Helmholtz formula, the optical force per unit length $\V F$ can be evaluated from a contour integral over the boundary,
\begin{equation}
\begin{aligned}
    \V F=\frac{a\epsilon_0}{4}(n_1^2-n_0^2)\left[\hat{\V x}\int_0^{2\pi}|\left(\phi_n^m\right)_{r=a}|^2\cos\theta\,d\theta\right.\\
    +\left.\hat{\V y}\int_0^{2\pi}|\left(\phi_n^m\right)_{r=a}|^2\sin\theta\,d\theta\right],
\end{aligned}
\end{equation}
where again $n_1$ and $n_0$ denote the refractive index of the core and cladding medium, respectively. Clearly, because of symmetry, the net force along $y$ is zero. In this regard, Eq. (21) is reduced to 
\begin{equation}
\begin{aligned}
    \V F=\hat{\V x}\frac{a\epsilon_0}{4}(n_1^2-n_0^2)\left[\int_0^{2\pi}\left(\sum_{k=n-1}^{k=n+1}|c_k^m G_{0,k}|^2\right)_{r=a}\cos\theta\,d\theta\right.\\
    +\left.\int_0^{2\pi}\left(\sum_{k\neq j}c_k^mc_j^m G_{0,k}G_{0,j}^*\right)_{r=a}\cos\theta\,d\theta\right].
\end{aligned}
\end{equation}
By substituting $c_n^m$ into Eq. (22), one finds (Appendix A)
\begin{equation}
    \V F=\hat{\V x}Q\sin{\left(\frac{2mn\pi}{N+1}\right)},
\end{equation}
where $Q$ is a proportionality constant that can be obtained from the overlap integrals. We would like to note that the forces in this system, as expressed by Eq. (23), cannot be directly obtained from Eq. (13) given that they vary considerably across the waveguide array. Yet, interestingly, Eq. (13) can be used to evaluate the coefficient $Q$ (Appendix B). In this respect, it turns out that $Q$ can be obtained from 
\begin{equation}
    Q=C\frac{\sin q\sin{(q/2)}}{N\sin(Nq)-(N-1)\sin{\left[(N+1)q\right]}-\sin q},
\end{equation}
where the constant $C$ is given by $C=-4[PL/(k_0 c)](\sqrt{2\Delta}/a^2)(U^2 W/V^3)[K_1 (WD/a)/K_1^2 (W)]$, $q=2m\pi/(N+1)$. Here, $V=k_0 a n_1\sqrt{2\Delta}$ is the waveguide $V$ number, $K_j (x)$ is a modified Bessel function of order $j$ and the quantities $U$ and $W$ are defined as $U=a\sqrt{k_0^2n_1^2-\beta^2}$, $W=a\sqrt{\beta^2-k_0^2n_2^2}$ \cite{snyder2012optical,okamoto2021fundamentals}. These results are now compared to the force distribution resulting in a linear waveguide array involving 8 single-mode elements. In this system, the core radius $a=5.3\mu m$ and the distance between elements is $D=20\mu m$. Moreover, $n_1=1.5$ while $n_0=n_1 (1-\Delta)$ where $\Delta$ is $2×10^{-3}$. In all cases, we assume that the power flowing in each supermode is $1\rm{W}$. From finite element computations (based on either the Maxwell stress tensor or the Minkowski-Helmholtz formula), we find that Eqs. (23, 24) provide a good description of the force distribution, with an error that is less than $7\%$. This error is attributed to the validity of the coupled-mode theory itself. Yet, given the complexity of the system, Eqs. (23, 24) do provide valuable information as to the stress variation across the lattice. 

As indicated before, optical forces also manifest themselves in scattering configurations. In this respect, the Minkowski-Helmholtz formula can be utilized to evaluate the optical force exerted on a dielectric scatterer such as a Mie particle. To demonstrate this aspect, we consider for simplicity a plane wave that is incident in vacuum on a particle of refractive index $n=3.5$ and radius $a=200nm$ i.e., $\V E=\hat{\V x}E_0  \exp(ik_0z)$, where $E_0=10^6 V/m$. Numerical simulations were carried out using finite element schemes over a range of wavelengths ($1\mu m-2\mu m$). In Fig. (6), we compare the optical force obtained via three different methods; the Minkowski-Helmholtz formula, the Maxwell stress tensor formalism \cite{jackson1999classical} and the analytic expression for the optical force on a Mie particle, given by $\V F=\left(1-\langle\cos\theta\rangle\right)\sigma_{sc}\langle\V S\rangle/c$ \cite{kerker2013scattering,bohren2008absorption,salandrino2012generalized}. In the latter expression,  $\langle\V S\rangle$ represents the time averaged Poynting vector, $\langle\cos\theta\rangle$ is the so-called average cosine function, $c$ is the speed of light and $\sigma_{sc}$ denotes the particle’s scattering cross section. As Fig. (6) reveals, these three procedures produce exactly the same results. We would like to emphasize that in deploying the Minkowski-Helmholtz formula, the numerical algorithm utilizes only the electric field vector on the surface while the Maxwell stress tensor approach involves in addition the magnetic field.
\begin{figure}
\includegraphics[width=8.6cm]{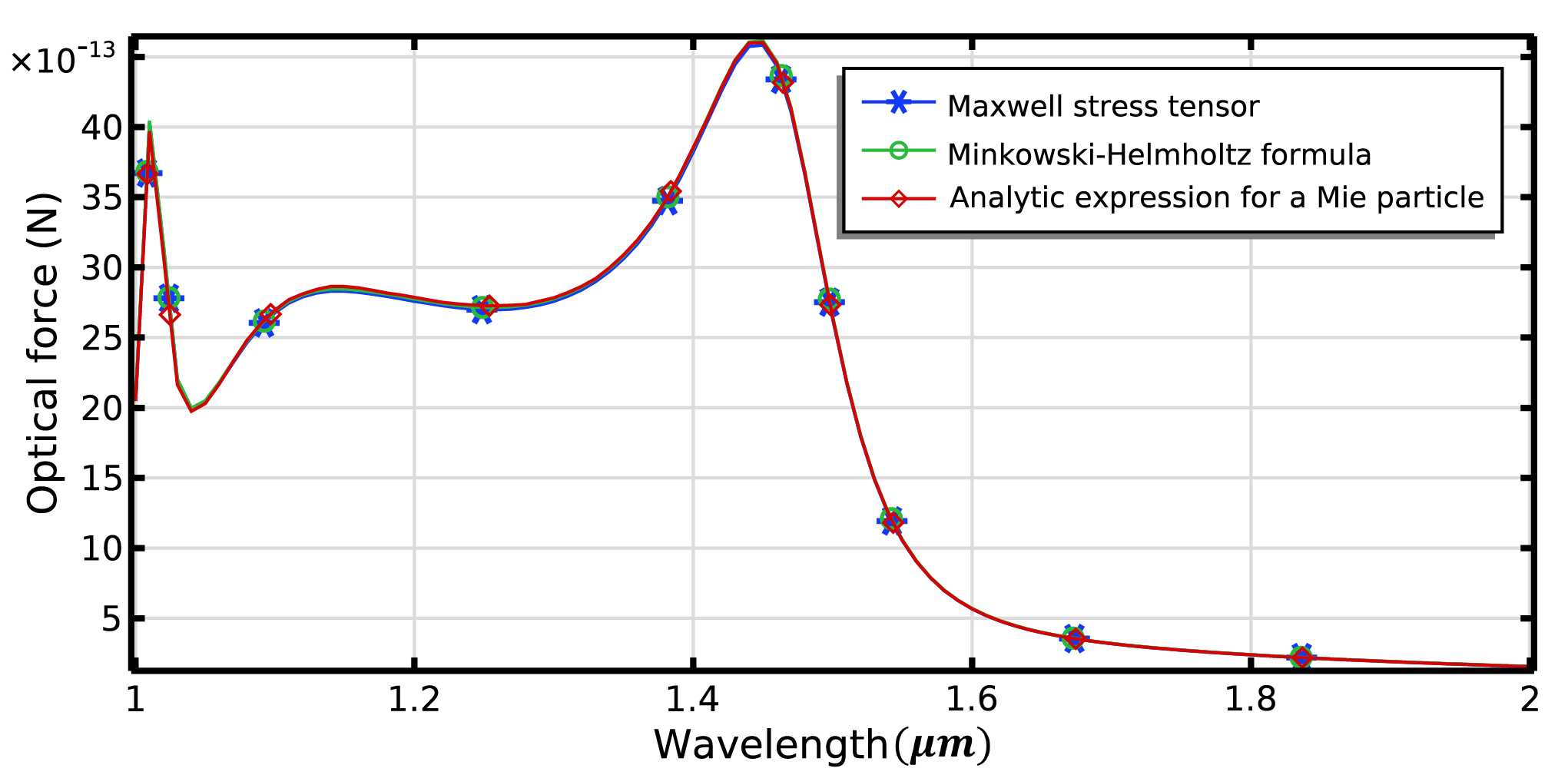}
\caption{\label{fig:6}Numerical simulations of the optical force exerted on a Mie particle. The optical force is calculated via the Maxwell stress tensor formalism (blue line), the Minkowski-Helmholtz formula (green line) as well as the analytic expression for a Mie particle (red line).}
\end{figure} 

As previously indicated, electromagnetic forces are also at play in settings where the refractive index changes gradually in space. For example, in a graded-index (GRIN) parabolic fiber of radius $a$, the refractive index varies with the radius $r$ according to $\epsilon_r(r)= n^2 (r)=n_1^2 (1-2\Delta(r/a)^2)$. In weakly guiding parabolic fibers, the dominant transverse electric field of an OAM-free Laguerre-Gauss mode $(LG_{lm})$ is given by $E_{lm}\propto \eta^l e^{-\eta^2/2} L_{m-1}^l (\eta^2 )\times\cos(l\theta+\psi)$ where $\eta=\sqrt{V}(r/a)$, $V=k_0n_1 a\sqrt{2\Delta}$, and $L_{m-1}^l (\eta^2)$ are generalized Laguerre polynomials \cite{snyder2012optical,okamoto2021fundamentals}. From Eq. (2), one can then readily obtain the Minkowski-Helmholtz force density $\langle\V {f_M}\rangle$ within the core region corresponding to various $LG_{lm}$ modes. The force densities associated with the $LG_{01}$ and $LG_{11}$ modes are depicted in Fig. 7. In all cases, they are pointing radially outwards. At this point, one may ask how the Minkowski-Helmholtz formula can deduce the result of Eq. (13) in this more complex arrangement. To address this issue, let us assume the radius of a fiber $a$ is adiabatically increased by $\delta a$. Notice that in this process, the displacement must vary in a self-similar manner so that the parabolic index profile is maintained. To satisfy this last condition, at each point the virtual displacement must vary according to $\delta\V\xi(\V r)=\hat{\V r}r\delta a/a$. In this case, the virtual work is
\begin{equation}
    \begin{aligned}
    \delta W&=L\int_0^{2\pi}d\theta\int_0^a rdr\V f(\V r)\cdot\delta\V \xi(\V r)\\
    &=\frac{2\pi L\delta a}{a}\int_0^a r^2dr\left(-\frac{1}{4}|\V E|^2\nabla\epsilon\right)\cdot\hat{\V r}.
    \end{aligned}
\end{equation}
To establish the equivalence of Eq. (25) to Eq. (13), we carried out numerical simulations in a weakly guiding GRIN parabolic fiber with $a=25 \mu m$, $n_1=1.5$ and $\Delta\approx1×10^{-3}$. The operating wavelength was taken to be $1\mu m$ and the total power is $P=1 \rm{W}$. Note that the effective refractive index $n_{eff}$ for the $LG_{lm}$ mode is given by  $n_{eff,lm}=n_1\sqrt{1-\left[2(2m+l-1)\sqrt{2\Delta}\right]/(k_0an_1)}$. In this case, for the same virtual enlargement $\delta a$, one finds that the ratio of virtual works (as obtained from Eq. (13) and Eq. (25)) is $1.0006$ and $1.0024$ for the $LG_{01}$ and $LG_{11}$ mode, respectively. This clearly shows that indeed Eqs. (13, 25) yield identical results. The small departure from unity is attributed to the non-vectorial paraxial treatment of the multi-mode parabolic waveguide. Finally, we would like to emphasize that while Eqs. (13, 25) are logistically equivalent under an adiabatic expansion, they are of little physical relevance. Physically, the structural deformation of a waveguide will be dictated by the Minkowski-Helmholtz force density $\langle\V {f_M}\rangle$ when taken in conjunction with the elastic properties of the material system \cite{timoshenko1970theory}. For example, from Fig. 7(b), one will expect from Eq. (2) that the actual fiber will be elliptically elongated, something that cannot be directly captured from Eq. (13). If the force density $\langle\V {f_M}\rangle$ is also coupled with the photoelastic properties of the underlying materials, it could also be useful for analyzing stimulated Brillouin scattering processes \cite{qiu2013stimulated}.

\begin{figure}
\includegraphics[width=8.6cm]{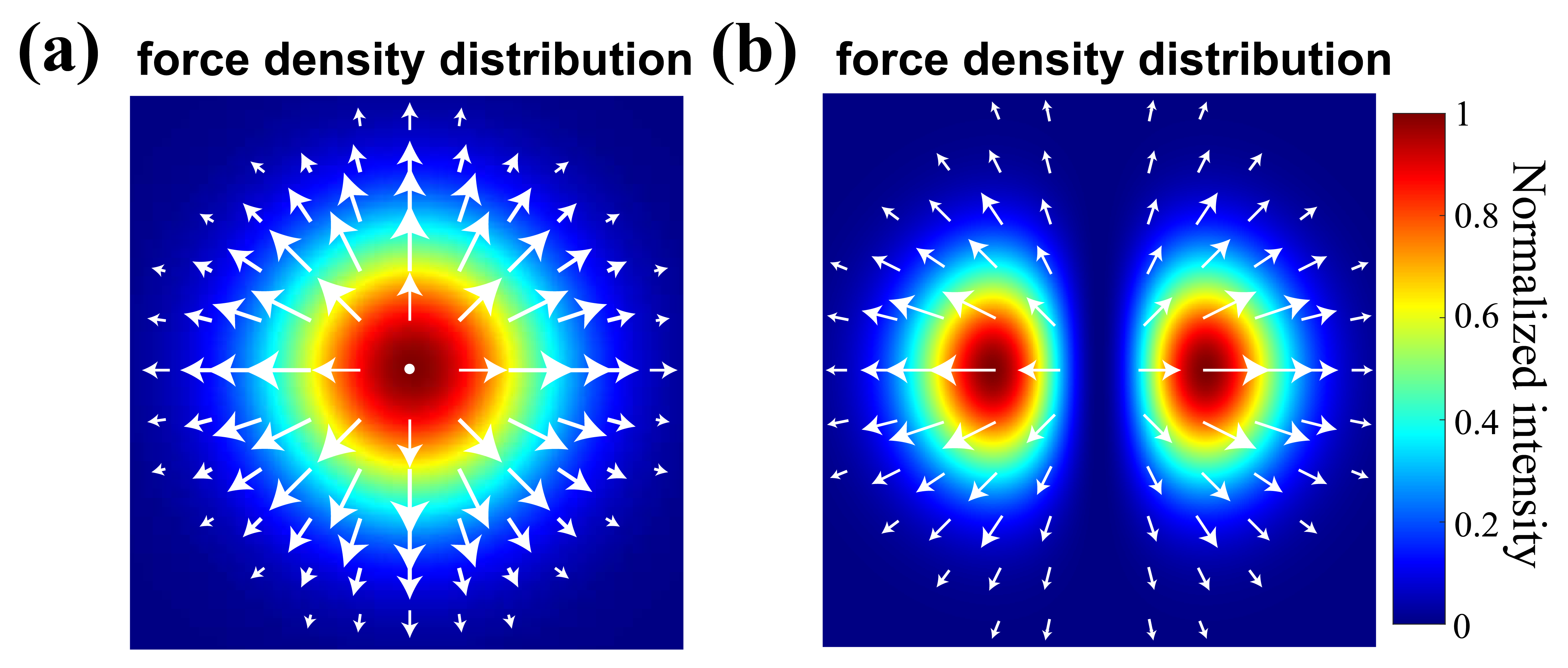}
\caption{\label{fig:7}Minkowski-Helmholtz force density $\langle\V {f_M}\rangle$ corresponding to (a) the $LG_{01}$ mode and (b) the $LG_{11}$ mode.}
\end{figure} 

\section{Conclusion}
In this work, we have presented a rigorous proof concerning the equivalence between energy-based methodologies and the Maxwell stress tensor formalism. This proof was based on the Minkowski-Helmholtz formula and the electromagnetic variation theorem as they apply in a lossless or Hermitian system. Our theoretical analysis is general and can be used in any arbitrary dielectric system involving elements like optical cavities and waveguides. In addition, we showed that the Minkowski-Helmholtz formula can provide an elegant way to compute optical forces emerging in a variety of diverse and complex arrangements. These include multi-element waveguide arrays, dielectric scatterers and graded-index waveguides. As indicated in our work, the Minkowski-Helmholtz formula not only offers a powerful intuitive tool in understanding optical forces but also provides a straightforward avenue in computing these forces in more involved settings where energetic approaches cannot account for optically induced internal stresses. Finally, it will be of interest to investigate how these concepts can be extended in the case of non-Hermitian configurations like those associated with parity-time symmetry that could in principle display exceptional points \cite{miri2019anomalous}.

\section*{ACKNOWLEDGMENTS}
 This work was partially supported by ONR MURI (N00014-20-1-2789), AFOSR MURI (FA9550-20-1-0322, FA9550-21-1-0202), DARPA (D18AP00058), Office of Naval Research (N00014-16-1-2640, N00014-18-1-2347, N00014-19-1-2052, N00014-20-1-2522, N00014-20-1-2789), National Science Foundation (NSF) (DMR-1420620, EECS-1711230, CBET 1805200, ECCS 2000538, ECCS 2011171), Air Force Office of Scientific Research (FA9550-14-1-0037,  FA9550-20-1-0322, FA9550-21-1-0202), MPS Simons collaboration (Simons grant 733682), W. M. Keck Foundation, USIsrael Binational Science Foundation (BSF: 2016381), US Air Force Research Laboratory (FA86511820019) and the Qatar National Research Fund (grant NPRP13S0121-200126). G.G.P. acknowledges the support of the Bodossaki Foundation.
\vspace*{-0.3cm}

\section*{Appendix A}
 We here derive Eq. (23) from Eq. (22). By substituting $c_n^m$ into Eq. (22), we obtain
\renewcommand{\theequation}{A.\arabic{equation}}

\setcounter{equation}{0}

\begin{equation}
\begin{aligned}
\V F&=\hat{\V x}\frac{a\epsilon_0}{4}(n_1^2-n_0^2)\frac{2}{N+1}\\
&\int_0^{2\pi}d\theta\cos\theta\left[\sum_{k=n-1}^{n+1}\sin^2\left(\frac{mk\pi}{N+1}\right)\left(|G_{0,k}|^2\right)_{r=a}\right.\\
&+\left.\sum_{k\neq j}\sin\left(\frac{mk\pi}{N+1}\right)\sin\left(\frac{mj\pi}{N+1}\right)\left(G_{0,k}G_{0,j}^*\right)_{r=a}\right]\\
\end{aligned}
\end{equation}
By keeping in mind that $G_0$ represents the local $LP_{01}$ mode profile in each lossless element $(G_0=G_0^*)$, and because of symmetry, we find
\begin{equation}
\begin{aligned}
  &\int_{0}^{2\pi}{\left(\left|G_{0,n}\right|^2\right)_{r=a}\cos{\theta}d\theta}=0,\\
  &\int_{0}^{2\pi}{\left(\left|G_{0,n-1}\right|^2\right)_{r=a}\cos{\theta}d\theta}\\
  &\qquad\qquad\qquad=-\int_{0}^{2\pi}{\left(\left|G_{0,n+1}\right|^2\right)_{r=a}\cos{\theta}d\theta},\\
  &\int_{0}^{2\pi}{\left(G_{0,n-1}G_{0,n}\right)_{r=a}\cos{\theta}d\theta}\\
  &\qquad\qquad\qquad=-\int_{0}^{2\pi}{\left(G_{0,n+1}G_{0,n}\right)_{r=a}\cos{\theta}d\theta},\\
  &\int_{0}^{2\pi}{\left(G_{0,n-1}G_{0,n+1}\right)_{r=a}\cos{\theta}d\theta}=0.
\end{aligned}  
\end{equation}
From here, Eq. (A1) can be reduced to

\begin{equation}
    \begin{aligned}
      \V{F} &= \hat{\V{x}}\frac{a\epsilon_0}{2\left(N+1\right)}\left(n_1^2-n_0^2\right)\left\{\int_{0}^{2\pi}\left\{\sin^2\left[\frac{m\left(n+1\right)\pi}{N+1}\right]\right.\right.\\
      & -\left.\sin^2{\left[\frac{m\left(n-1\right)\pi}{N+1}\right]}\right\}\left(\left|G_{0,n+1}\right|^2\right)_{r=a}\cos{\theta}d\theta\\
      & +\int_{0}^{2\pi}2\sin{\left(\frac{mn\pi}{N+1}\right)}\left\{\sin{\left[\frac{m\left(n+1\right)\pi}{N+1}\right]}\right.\\
      & -\left.\left.\sin{\left[\frac{m\left(n-1\right)\pi}{N+1}\right]}\right\}\left(G_{0,n}G_{0,n+1}\right)_{r=a}\cos{\theta}d\theta\right\}\\
      &=\hat{\V{x}}\frac{a\epsilon_0}{2\left(N+1\right)}\left(n_1^2-n_0^2\right)\left[\sin{\left(\frac{2mn\pi}{N+1}\right)}\sin{\left(\frac{2m\pi}{N+1}\right)}\right.\\
      &\int_{0}^{2\pi}\left(\left|G_{0,n+1}\right|^2\right)_{r=a}\cos{\theta}d\theta+2\sin\left(\frac{2mn\pi}{N+1}\right)\\
      &\left.\sin\left(\frac{m\pi}{N+1}\right)\int_{0}^{2\pi}{\left(G_{0,n}G_{0,n+1}\right)_{r=a}\cos{\theta}d\theta\ }\right]\\
      &=\hat{\V{x}}\frac{a\epsilon_0}{N+1}\left(n_1^2-n_0^2\right)\sin{\left(\frac{2mn\pi}{N+1}\right)}\sin{\left(\frac{m\pi}{N+1}\right)}\\
      &\left[\cos{\left(\frac{m\pi}{N+1}\right)}\int_{0}^{2\pi}{\left(\left|G_{0,n+1}\right|^2\right)_{r=a}\cos{\theta}d\theta}\right.\\
      &+\left.\int_{0}^{2\pi}{\left(G_{0,n}G_{0,n+1}\right)_{r=a}\cos{\theta}d\theta\ }\right].\ \ \ 
    \end{aligned}
\end{equation}
By introducing the quantity
\begin{equation}
    \begin{aligned}
    Q &=\frac{a\epsilon_0}{N+1}\left(n_1^2-n_0^2\right)\sin{\left(\frac{m\pi}{N+1}\right)}\\
    &\left[\cos{\left(\frac{m\pi}{N+1}\right)}\int_{0}^{2\pi}{\left(\left|G_{0,n+1}\right|^2\right)_{r=a}\cos{\theta}d\theta}\right.\\
    &+\left.\int_{0}^{2\pi}{\left(G_{0,n}G_{0,n+1}\right)_{r=a}\cos{\theta}d\theta\ }\right],
    \end{aligned}
\end{equation}
Eq. (A3) can now be rewritten as
\begin{equation}
    \V{F}=\hat{\V{x}}Q\sin{\left(\frac{2mn\pi}{N+1}\right)}.
\end{equation}
\section*{Appendix B}
In this section, we will show that Eq. (13) can be utilized to evaluate the overlap integrals in Eq, (A5), and hence the quantity $Q$. From coupled mode theory, the propagation constant of each supermode is given by \cite{snyder2012optical}
\begin{equation}
    \beta_m=\beta_0+2\kappa\cos{\left(\frac{m\pi}{N+1}\right)},
\end{equation}
where $m=1,2,3,\ldots,N$ denotes the supermode index. As indicated in the main text, each waveguide element is here assumed to be cylindrical (of radius $a$) and single-moded, i.e., supporting only the $LP_{01}$ mode. The distance between core centers is $D$. In this case, the coupling strength between successive elements is given by 
\begin{equation}
    \kappa=\frac{\sqrt{2\Delta}}{a}\frac{U^2}{V^3}\frac{K_0\left(WD/a\right)}{K_1^2\left(W\right)},
\end{equation}
where $\Delta=(n_1-n_0)/n_1$ is the normalized waveguide index difference, $V=k_0an_1\sqrt{2\Delta}$ is the $V$ number and $K_j\left(x\right)$ is a modified Bessel function of order $j$. The quantities $U$ and $W$ are defined as $U=a\sqrt{k_0^2n_1^2-\beta^2}$, $W=a\sqrt{\beta^2-k_0^2n_2^2}$, and can be determined from the eigenvalue equation $UJ_1(U)/J_0(U)=WK_1(W)/K_0(W)$ \cite{snyder2012optical,okamoto2021fundamentals}. From Eq. (A6), one can obtain the effective index $n_{eff}$ of each supermode, i.e., $n_{eff,m}=\beta_m/k_0$. If the distance between successive elements is virtually altered by $\delta D$, then the virtual work performed can be obtained from Eq. (12), that is $\delta W=\left(PL/c\right)\left(dn_{eff}/dD\right)\delta D=[2PL/\left(k_0c\right)]\cos[m\pi/(N+1)]\left(d\kappa/dD\right)\delta D$. From here, one finds that
\begin{figure}
\includegraphics[width=8.6cm]{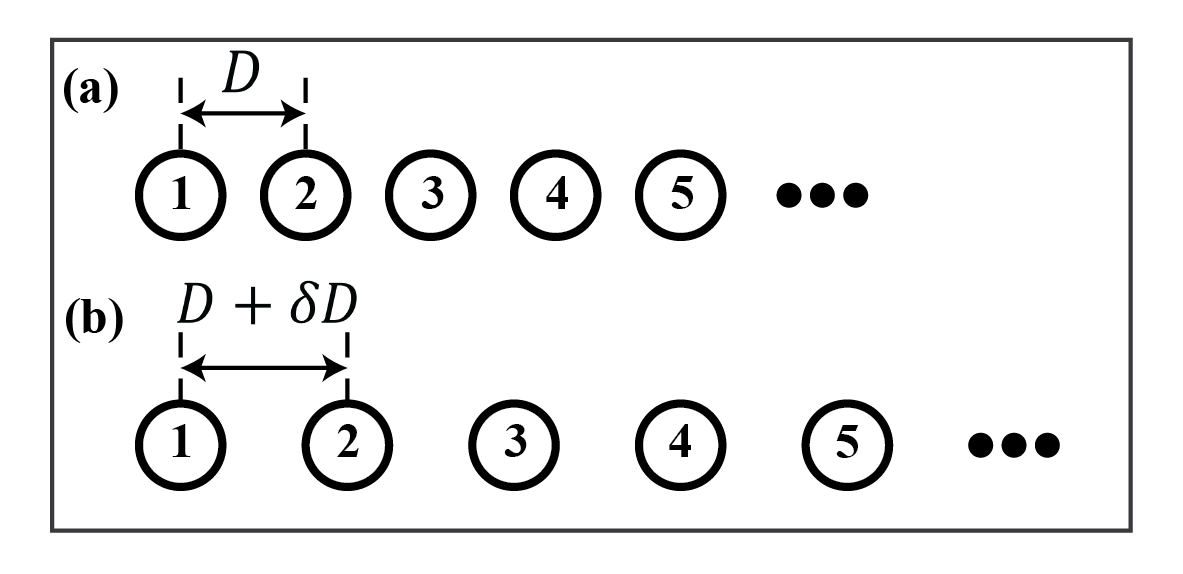}
\caption{\label{fig:A1}A uniform virtual expansion of a linear waveguide array. (a) and (b) represent the structure before and after the virtual displacement $\delta D$.}
\end{figure} 
\begin{equation}
    \delta W=-2\frac{PL}{k_0c}\cos{\left(\frac{m\pi}{N+1}\right)}\frac{\sqrt{2\mathrm{\Delta}}}{a^2}\frac{U^2W}{V^3}\frac{K_1\left(WD/a\right)}{K_1^2\left(W\right)}\delta D.
\end{equation}
If we assume the first element is kept fixed while the distance between adjacent elements changes from $D$ to $D+\delta D$ (shown as in Fig. (8)), each waveguide will be displaced by $\delta\V{\xi}_n=\hat{\V{x}}\left(n-1\right)\delta D$. In this respect, the work produced by the system is given by
\begin{equation}
    \delta W=\sum_{n}{\V{F}_n\cdot\delta\V{\xi}_n}.
\end{equation}
By substituting Eq. (A5) into Eq. (A9), one finds
\begin{equation}
    \begin{aligned}
    \delta W &=\sum_{n=1}^{N}{\hat{\V{x}}Q\sin{\left(\frac{2mn\pi}{N+1}\right)}\cdot\hat{\V{x}}\left(n-1\right)\delta D}\\
    &=Q\delta D\sum_{n=1}^{N}{\sin{\left(\frac{2mn\pi}{N+1}\right)}\left(n-1\right)}.
    \end{aligned}
\end{equation}
Given that
\begin{equation}
    \sum_{k=1}^{n}\sin{\left(kx\right)}=\frac{\sin{\left[(n+1)x/2\right]\sin{\left(nx/2\right)}}}{\sin{\left(x/2\right)}},
\end{equation}
and
\begin{equation}
    \sum_{k=1}^{n}{k\sin{\left(kx\right)}}=\frac{\sin{\left(nx\right)}}{4\sin^2{\left(x/2\right)}}-\frac{n\cos{\left[\left(2n+1\right)x/2\right]}}{2\sin{\left(x/2\right)}},
\end{equation}
Eq. (A10) can now be rewritten as
\begin{equation}
    \begin{aligned}
    \delta W&=Q\delta D\left\{\frac{\sin{\left(Nq\right)}}{4\sin^2{\left(q/2\right)}}-\frac{N\cos{\left[\left(2N+1\right)q/2\right]}}{2\sin{\left(q/2\right)}}\right.\\
    &\left.-\frac{\sin{\left[(N+1)q/2\right]\sin{\left(Nq/2\right)}}}{\sin{\left(q/2\right)}}\right\}\\
    &=Q\delta D\left\{\frac{\sin{\left(Nq\right)}-\sin{q}}{4\sin^2{\left(q/2\right)}}-\frac{\left(N-1\right)\cos{\left[\left(2N+1\right)q/2\right]}}{2\sin{\left(q/2\right)}}\right\}\\
    &=Q\delta D\left\{\frac{N\sin{\left(Nq\right)}-\left(N-1\right)\sin{\left[\left(N+1\right)q\right]}-\sin{q}}{4\sin^2{\left(q/2\right)}}\right\},
    \end{aligned}
\end{equation}
where $q=2m\pi/(N+1)$. By combining Eqs. (A8), and (A13), one obtains
\begin{equation}
    \begin{aligned}
    &Q\left\{\frac{N\sin{\left(Nq\right)}-\left(N-1\right)\sin{\left[\left(N+1\right)q\right]}-\sin{q}}{4\sin^2{\left(q/2\right)}}\right\}\\
    &=-2\frac{PL}{k_0c}\cos{\left(\frac{q}{2}\right)}\frac{\sqrt{2\mathrm{\Delta}}}{a^2}\frac{U^2W}{V^3}\frac{K_1\left(WD/a\right)}{K_1^2\left(W\right)},
    \end{aligned}
\end{equation}
from where we can determine the value of $Q$ (Eqs. (23, 24))
 \begin{equation}
     \begin{aligned}
      Q&=-4\frac{PL}{k_0c}\frac{\sqrt{2\mathrm{\Delta}}}{a^2}\frac{U^2W}{V^3}\frac{K_1\left(WD/a\right)}{K_1^2\left(W\right)}\\
      &\left\{\frac{\sin{q}\sin{\left(q/2\right)}}{N\sin{\left(Nq\right)}-\left(N-1\right)\sin{\left[\left(N+1\right)q\right]}-\sin{q}}\right\}.
     \end{aligned}
 \end{equation}


\nocite{*}


%

\end{document}